\pdfoutput=1

\documentclass[11pt,twoside,a4paper,cmspaper,final,collab]{cms-tdr}
\def\svnVersion{64758}\def\cmsCernNo{2011-034}\def\cmsCernDate{\today}\def\cmsMessage{Submitted to Physical Review Letters}

\begin{document}\cmsNoteHeader{BPH-10-005}

\hyphenation{had-ron-i-za-tion}
\hyphenation{cal-or-i-me-ter}
\hyphenation{de-vices}
\RCS$Revision: 50371 $
\RCS$HeadURL: svn+ssh://alverson@svn.cern.ch/reps/tdr2/papers/BPH-10-005/trunk/BPH-10-005.tex $
\RCS$Id: BPH-10-005.tex 50371 2011-04-14 19:52:44Z alverson $

\def\invnb   {\ensuremath{\mbox{\,nb}^{-1}}}
\def\invpb   {\ensuremath{\mbox{\,pb}^{-1}}}
\def\nb   {\ensuremath{\,\mbox{nb}}}
\def\ub   {\ensuremath{\,\mu\mbox{b}}}
\def\um   {\ensuremath{\,\mu\mbox{m}}}
\def\psik {\ensuremath{B^+\to \JPsi K^+}}
\def\psiks {\ensuremath{B^0\to \JPsi \KS}}
\def\psikst {\ensuremath{B^0\to \JPsi K^{*0}}}
\def\psiphi {\ensuremath{B^0_s\to \JPsi \phi}}
\def\tev  {\ensuremath{\mbox{\,TeV}}}
\def\tmu   {\ensuremath{\tt trackerMuon}}
\def\tmus   {\ensuremath{\tt trackerMuons}}
\def\cc   {\ensuremath{c\bar{c}}}
\def\ppMuX {{\tt ppMuX}}
\def\pt {\ensuremath{p_T}}
\def\ra                 {\ensuremath{\rightarrow}\xspace}
\def\Bz      {\ensuremath{B^0}\xspace}
\def\Bd      {\ensuremath{B^0_d}\xspace}
\def\Bp      {\ensuremath{B^+}\xspace}
\def\Bs      {\ensuremath{B^0_s}\xspace}
\def\KS    {\ensuremath{K^0_{\scriptscriptstyle S}}\xspace}
\newcommand\dbline{\noalign{\vskip 0.10truecm\hrule}\noalign{\vskip 2pt}\noalign{\hrule\vskip 0.10truecm}}
\newcommand\sgline{\noalign{\vskip 0.10truecm\hrule\vskip 0.10truecm}}
\newcommand{\msp}{\ensuremath{\phantom{-}}}
\newcommand{\lum} {\ensuremath{\mathcal{L}}\xspace}
\newcommand{\chisq}{\ensuremath{\chi^2}\xspace}
\def\ctau {\ensuremath{c\times\tau}\xspace}
\def\MB   {\ensuremath{m_{\PB}}}
\def\MBz  {\ensuremath{m_{\PBz}}}
\def\Bbar    {\kern 0.18em\overline{\kern -0.18em B}{}\xspace}
\def\BB      {\ensuremath{B\Bbar}\xspace}
\def\calL{{\ensuremath{\cal L}}\xspace}
\newcommand{\calB}{\ensuremath{{\cal B}}}
\def\dsdpt {\ensuremath{d\sigma/dp_\mathrm{T}^{\PB}}}
\def\dsdy {\ensuremath{d\sigma/dy^{\PB}}}
\def\ptb {\ensuremath{p_\mathrm{T}^{\PB}}}
\def\pt {\ensuremath{p_\mathrm{T}}}
\def\yb {\ensuremath{\left|y^{\PB}\right|}}
\def\psiks {{\ensuremath{\JPsi\, \PKzS}}}
\def\Bpsiks {{\ensuremath{\PBz\to\psiks}}}
\def\psimumu {\ensuremath{\JPsi\to\Pgmp\Pgmm}}
\def\kspipi {\ensuremath{\PKzS\to\Pgpp\Pgpm}}
\def\mev    {\ensuremath{\MeV}}
\def\pp {\ensuremath{\Pp\Pp}}
\newcommand\T{\rule{0pt}{2.6ex}}
\newcommand\B{\rule[-1.2ex]{0pt}{0pt}}
\providecommand{\cPqb}{\ensuremath{\mathrm{b}}} 
\providecommand{\cPaqb}{\ensuremath{\overline{\mathrm{b}}}} 
\def\totalsigma{\ensuremath{33.2\pm 2.5\pm 3.5~\mu\textrm{b}}}

\cmsNoteHeader{BPH-10-005} 
\title{Measurement of the \PBz\ Production Cross Section in \pp\ Collisions at $\sqrt{s} = 7\TeV$}

\ifthenelse{\boolean{cms@external}}{%
\author{The CMS Collaboration}\affiliation{}
}
{%
}

\date{\today}

\abstract{
Measurements of the differential production cross sections $\dsdpt$ and $\dsdy$ for $\PBz$
mesons produced in \pp\ collisions at $\sqrt{s} = 7\TeV$ are presented.
The dataset used was collected by the
CMS experiment at the LHC and corresponds to
an integrated luminosity of $40\invpb$.
The production cross section is measured from
\PBz\ meson decays
reconstructed in
the exclusive final state \psiks, with the subsequent decays \psimumu\ and \kspipi.
The total
cross section for $\ptb > 5\GeV$ and $\yb < 2.2$ is measured to be
\totalsigma, where the first uncertainty is statistical and the second is
systematic.}

\hypersetup{%
pdfauthor={CMS Collaboration},%
pdftitle={Measurement of the B0 production cross section measurements in pp Collisions at sqrt(s) = 7 TeV},%
pdfsubject={CMS},%
pdfkeywords={CMS, B physics}}

\maketitle 

Cross sections for heavy quark production in hard scattering interactions
have been studied at $\Pp\Pap$ colliders
at center-of-mass energies from 630 \GeV~\cite{Albajar:1988th} to 1.96 \TeV
~\cite{Abe:1995dv,Abulencia:2006ps,Abachi:1994kj} and in $\Pp-$nucleus collisons with
beam energies from 800 to 920 \GeV~\cite{Zaitsev:2009zz}. 
The expected cross sections can
be calculated in perturbative Quantum Chromodynamics. The comparison between data and 
predictions provides a critical test of next-to-leading order (NLO) calculations~\cite{NDE}.
Considerable progress has been achieved in understanding heavy quark 
production at Tevatron energies, largely resolving earlier discrepancies~\cite{Cacciari04}, 
but substantial theoretical 
uncertainties remain due to the dependence on the renormalization and 
factorization scales. Measurements of \cPqb-hadron 
production at 7 \TeV\ provided by the Large Hadron Collider 
(LHC)~\cite{BPH-10-007,BPH-10-004,LHCb} represent 
a test at a new center-of-mass energy of theoretical approaches that aim to describe
heavy flavor production~\cite{Cacciari98,Kniehl08}.

This Letter presents the first measurement of the \PBz\ cross section
in \pp\ collisions at $\sqrt{s}=7\TeV$. Events with \PBz\ mesons
reconstructed from their decays to the final state \psiks, with \psimumu\ and 
\kspipi, are used to measure
\dsdpt, \dsdy, and the integrated cross section for 
transverse momentum \ptb\ $> 5 \gev$ and rapidity \yb\ $< 2.2$, where
$y$ is defined as $\frac{1}{2}\ln{\frac{E+p_L}{E-p_L}}$, $E$ is the particle energy, 
and $p_L$ is the
particle momentum along the counterclockwise beam direction.
As the \PBz\ and \PaBz\ are indistinguishable in this analysis, both mesons
are referred to as \PBz\ for the purposes of reconstruction and the final
results are divided by two to obtain an average.

The data sample collected by the Compact Muon Solenoid (CMS) detector at the LHC
corresponds to an integrated luminosity of
$39.6\pm1.6 \invpb$ and represents the entire 2010 dataset. 
A detailed description of the detector may be found 
elsewhere~\cite{JINST}. The main detector components used in this 
analysis are the silicon tracker and the muon systems. 

The silicon tracker measures charged particles within the pseudorapidity 
range $|\eta| < 2.5$, where $\eta = -\ln{\tan{(\frac{\theta}{2})}}$ and $\theta$ is the 
polar angle of the track relative to the counterclockwise beam direction.
It consists of 1440 silicon pixel and 15\,148 silicon strip detector modules and 
is located in the 3.8~T field of the superconducting solenoid. It provides an impact 
parameter resolution of $\sim$\,15~$\mu$m and a $\pt$ 
resolution of about 1.5\% for particles with transverse momenta up to
$100\GeV$. Muons are measured in the pseudorapidity range $|\eta|< 2.4$, 
with detection planes made using three technologies: drift tubes, cathode strip chambers, 
and resistive plate chambers.

Events are selected by a trigger requiring two muons without any
explicit requirement on the muon momentum. The muon candidates are fully
reconstructed offline, combining information from the silicon tracker and muon
detectors, and are required to be within the following kinematic acceptance
region: $\pt^{\Pgm}>3.3\GeV$ for $\left | \eta^{\Pgm}\right | < 1.3$; 
total momentum $p^{\Pgm}>2.9\GeV$ for $1.3 < \left | \eta^{\Pgm}\right | < 2.2$; and
$\pt^{\Pgm}>0.8\GeV$ for $2.2 < \left | \eta^{\Pgm}\right | < 2.4$.
Opposite-sign muon pairs are fit to a common vertex to form
\JPsi\ candidates, which are
required to be within $150\MeV$ of the world-average \JPsi\ mass~\cite{PDG2010}.

The \PKzS\ candidates are formed by fitting oppositely charged
tracks reconstructed with the CMS tracking algorithm~\cite{TRK_10_001} to a common 
vertex. Each track
is required to have at least 6 hits in the silicon tracker, a normalized
$\chi^2 < 5$, and a transverse impact parameter with respect to the luminous region
greater than 0.5 times its uncertainty. The reconstructed \PKzS\ decay vertex
must have a normalized $\chi^2 < 7$ and a transverse separation from the
luminous region at least $5$ times larger than the uncertainty on the separation. 
The \Pgpp\Pgpm\ invariant mass $m_{\PKzS}$ is required to
satisfy $478 < m_{\PKzS} < 518 \mev$, and the reconstructed mass
distribution is found to be in good agreement with the world-average value~\cite{PDG2010}.

The \PBz\ candidates are formed by combining a \JPsi\ candidate with a 
\PKzS\ candidate. 
A kinematic fit is performed with the two muons and the \PKzS\ candidate, in which the
invariant masses of the \JPsi\ and \PKzS\ candidates are constrained to their 
world-average values~\cite{PDG2010}. The \PBz\ vertex fit confidence level is
required to be greater than $1\%$ and the reconstructed $\PBz$ mass \MB\ must satisfy 
$4.9<\MB<5.7\gev$.  
When more than one candidate in a single 
event passes all the selection criteria, only the candidate with the highest 
\PBz\ vertex fit confidence level is retained, which results in the correct choice $99\%$
of the time in simulated events containing a true signal candidate. 
A total of 23\,174 \PBz\ candidates pass all selection criteria.

The efficiency of the \PBz\ reconstruction
is computed with a combination of techniques using the data and  
large samples of fully simulated signal events generated by 
\PYTHIA\ 6.422~\cite{PYTHIA}, decayed by \EVTGEN~\cite{EvtGen}, and simulated by 
\GEANTfour~\cite{GEANT4}. The trigger and 
muon-reconstruction efficiencies are obtained from a large sample of inclusive 
$\psimumu$ decays in data using a technique similar to that described 
in Ref.~\cite{BPH-10-002}, where one muon is identified with stringent quality requirements, 
and the second muon is identified using information either exclusively from the 
tracker (to measure the trigger and muon-identification efficiencies), or from the 
muon system (to measure the silicon tracking efficiency). 
Since the dimuon efficiencies are calculated as the product of the measured single muon
efficiencies, a correction ($1$--$6\%$), obtained from the simulation, is applied to take into account
efficiency correlations between the two muons.
The probabilities for the
muons to lie within the kinematic acceptance region and for the \PBz\ and 
\PKzS\ candidates to pass the selection requirements are determined from the simulated events.
To minimize the effect of the \PYTHIA\ modeling of the \ptb\ and \yb\ distributions 
on the efficiency calculation, the simulated events are reweighted to match the 
kinematic distributions observed in the data. The efficiencies for hadron-track 
reconstruction~\cite{TRK-10-002}, \PKzS\ reconstruction~\cite{QCD-10-007},
and for fulfilling the vertex quality requirement are found to be consistent between data and 
simulation within the available precision (up to $5\%$).

The total efficiency of this selection, defined as the fraction of $\Bpsiks$ decays
produced with $\ptb>5\GeV$ and $\yb<2.2$ that pass all criteria, is $1.7\%$. The efficiency 
ranges from $0.7\%$ 
for $\ptb\sim 5\GeV$, to $11.4\%$ for $\ptb>24\GeV$, with roughly equal losses due to
the dimuon kinematic acceptance, the dimuon trigger, and the \PKzS\ reconstruction.

The proper decay length of each selected $\PBz$ candidate is calculated as
$ct = (\MB/\ptb)L_{xy}$, where the transverse decay length $L_{xy}$ is the 
vector $\vec{s}$ pointing from the primary vertex~\cite{TRK_10_001} to the 
\PBz\ vertex projected onto the \PBz\ transverse momentum vector: 
$L_{xy} = (\vec{s}\cdot\vec{p}_{\textrm T}^{\PB})/{\left | \vec{p}_{\textrm T}^\PB\right |}$.

\begin{figure}[!h]
  \begin{center}
   \includegraphics[angle=90,width=0.49\textwidth]{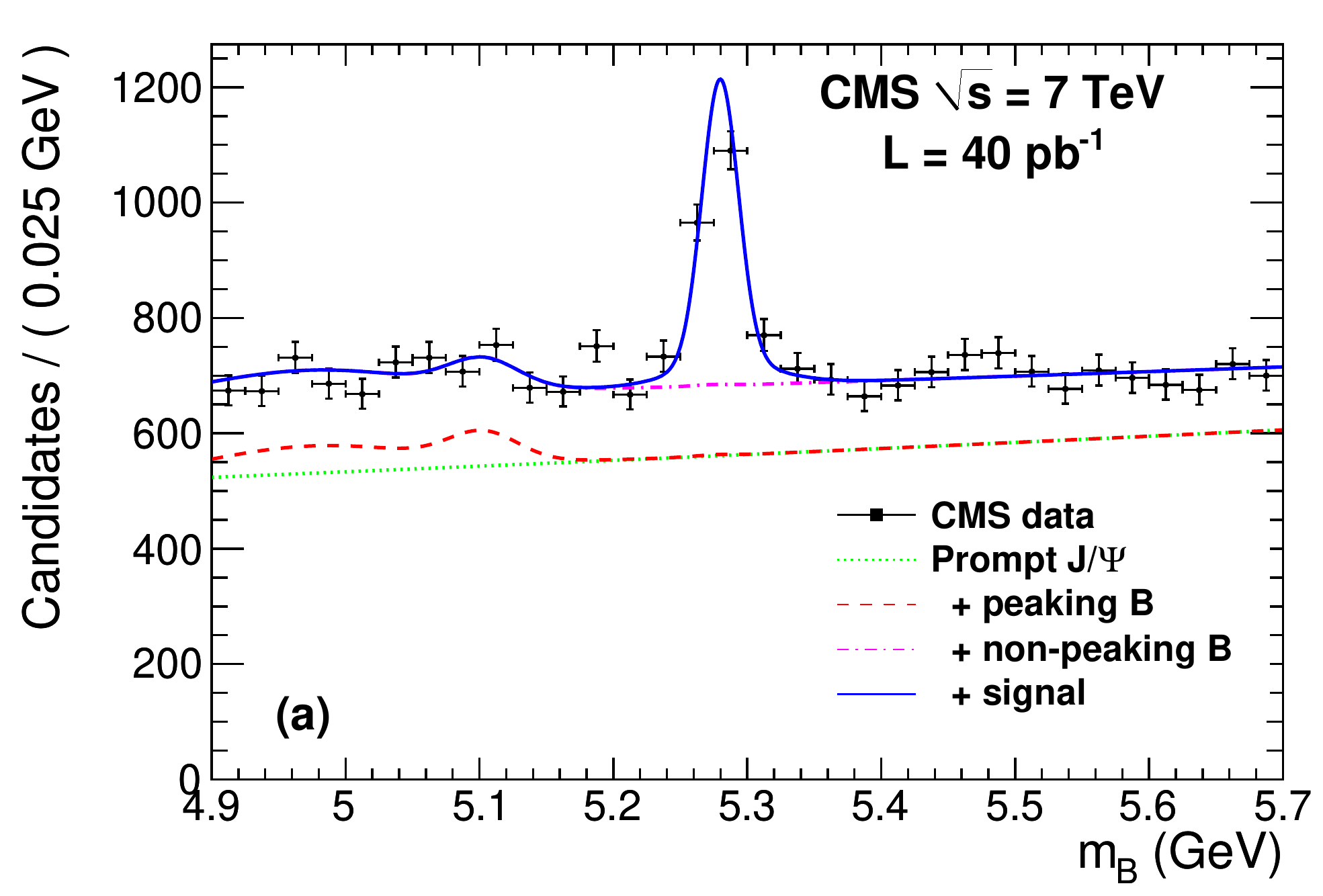}
   \includegraphics[angle=90,width=0.49\textwidth]{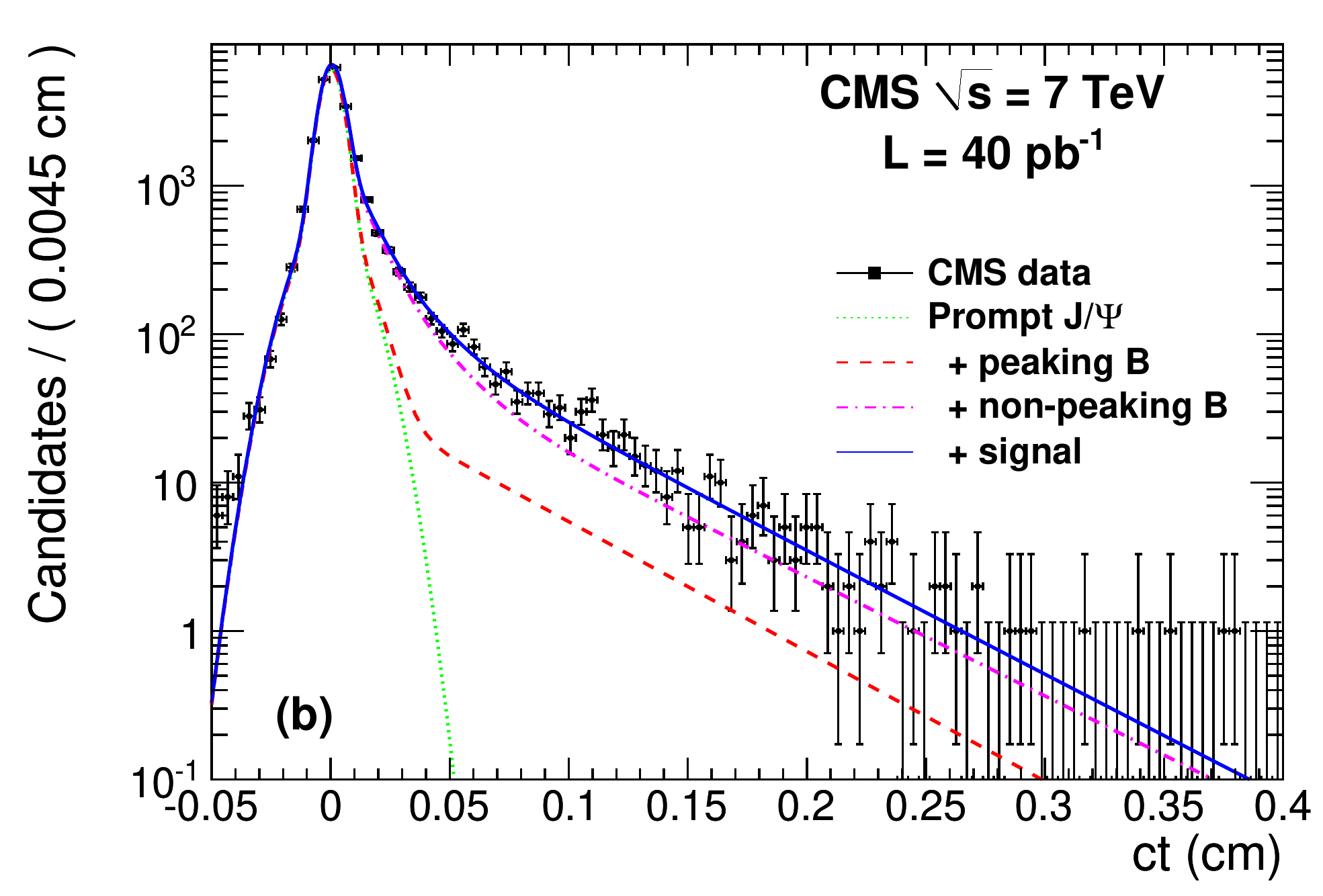}
    \caption{Projections of the fit results in (a) $\MB$ and (b) $ct$ for 
    $\ptb>5\GeV$ and $\yb < 2.2$.  The curves in
    each plot are as follows: the sum of all contributions (blue solid line);
    the prompt $\JPsi$ (green dotted); the sum of the prompt $\JPsi$ and peaking background (red dashed),
    and the sum of all backgrounds (purple dot-dashed).}
    \label{fig:fit}
  \end{center}
\end{figure}

Backgrounds are dominated by prompt and non-prompt $\JPsi$ production, with non-prompt
contributions from sources peaking and nonpeaking in \MB,
as shown in Fig.~\ref{fig:fit}.
In particular, misreconstructed \cPqb-hadron decays to final states with a $\JPsi$,
such as $\PB\to\JPsi\PKst$,
produce a broadly peaking structure in the region $\MB<5.2\GeV$.  A study of the 
dimuon invariant mass distribution
confirms that the contamination 
from events containing a misidentified $\JPsi$ is negligible after all selection criteria have been 
applied.

The signal yields in each $\ptb$ and $\yb$ bin
are obtained using an unbinned extended maximum-likelihood fit to $\MB$ and $ct$.  
The likelihood for event $j$ is obtained by summing the product of yield $n_i$ and 
probability density ${\cal P}_i$ for each of the signal and background hypotheses 
$i$.  Four individual components are considered: signal events, prompt $\JPsi$ events,
non-prompt $\cPqb\to\JPsi$ events that peak in $\MB$ (peaking), and non-prompt $\cPqb\to\JPsi$ 
events that do not peak in $\MB$ (nonpeaking).  
The extended likelihood function 
is the product of likelihoods for all events:
\begin{equation}
{\cal L} = \exp\left ( -\sum_{i=1}^{4} n_i \right ) \prod_j \left [ 
\sum_{i=1}^4 n_i{\cal P}_i(\MB;\, \vec{\alpha}_i){\cal P}_i(ct;\, \vec{\beta}_i) \right ].
\end{equation}
The probability density functions (PDFs), ${\cal P}_i$,
with shape parameters $\vec{\alpha}_i$ for $\MB$ and 
$\vec{\beta}_i$ for $ct$, are evaluated separately for each of the $i$ fit components.  
The yields $n_i$ are determined by maximizing $\cal L$ with 
respect to the yields and a subset of the PDF parameters. 

The PDF shapes are described below with the parameters obtained from data when possible. 
The $\MB$ PDFs are as follows: the sum of two 
Gaussian functions for the signal; exponential functions for the prompt and nonpeaking
backgrounds; 
and a sum of three Gaussian functions for the peaking 
background.  The resolution on $\MB$ for correctly reconstructed 
signal events from simulation is approximately $20\MeV$.
The $ct$ PDFs are as follows: a single exponential function convolved with the resolution 
function to describe the signal and peaking background components, where 
the lifetimes are allowed to be different; the sum of two exponential functions 
convolved with the resolution function for the nonpeaking component; and the 
pure resolution function for the prompt $\JPsi$ component. The resolution function,
a sum of two Gaussian functions, is common for signal and background and is measured in
data to have an average resolution of $71~\mu$m. 

The fit proceeds in several steps such that all background shapes are obtained
directly from data, except for the peaking component which is taken from simulation,
as are the signal \MB\ shapes.
This technique relies on the 
assumption that in the region $5.4 < \MB < 5.7\GeV$ (sideband) there 
are only two contributions: prompt $\JPsi$ and nonpeaking background.  
To obtain the
effective lifetime distribution of the nonpeaking background, the $\MB$ and $ct$ 
distributions in the $\MB$ sideband region are fit 
simultaneously for 
events in the inclusive $\PBz$ sample defined by $\ptb>5\GeV$ and $\yb < 2.2$.
In the second step, the signal $\PBz$ lifetime in the inclusive sample 
is determined by fitting $ct$ and $\MB$ simultaneously in the full $\MB$ range.  The result, 
$c\tau = 479\pm 22\,\mu$m (statistical uncertainty only), is in agreement with the 
world-average value, $457\pm 3\,\mu$m~\cite{PDG2010}.  
With the effective lifetimes for signal and non-prompt 
background fixed, the signal and background yields are fit in each bin of $\ptb$ and $\yb$, 
together with the parameters 
describing the $ct$ resolution and the shapes of the prompt and nonpeaking 
components in $\MB$.

The accuracy and robustness of the fit strategy were demonstrated by performing a large
set of pseudo-experiments, with each one 
corresponding to the yields observed in data, where signal and background 
events were generated randomly from the PDFs in each bin. No significant biases were observed
on the yields, and the statistical precision of the test was taken as 
the systematic uncertainty due to potential biases in the fit method. 
The fit uncertainties were also observed to be estimated properly.

\begin{figure}[!h]
\begin{center}
\includegraphics[width=0.49\textwidth]{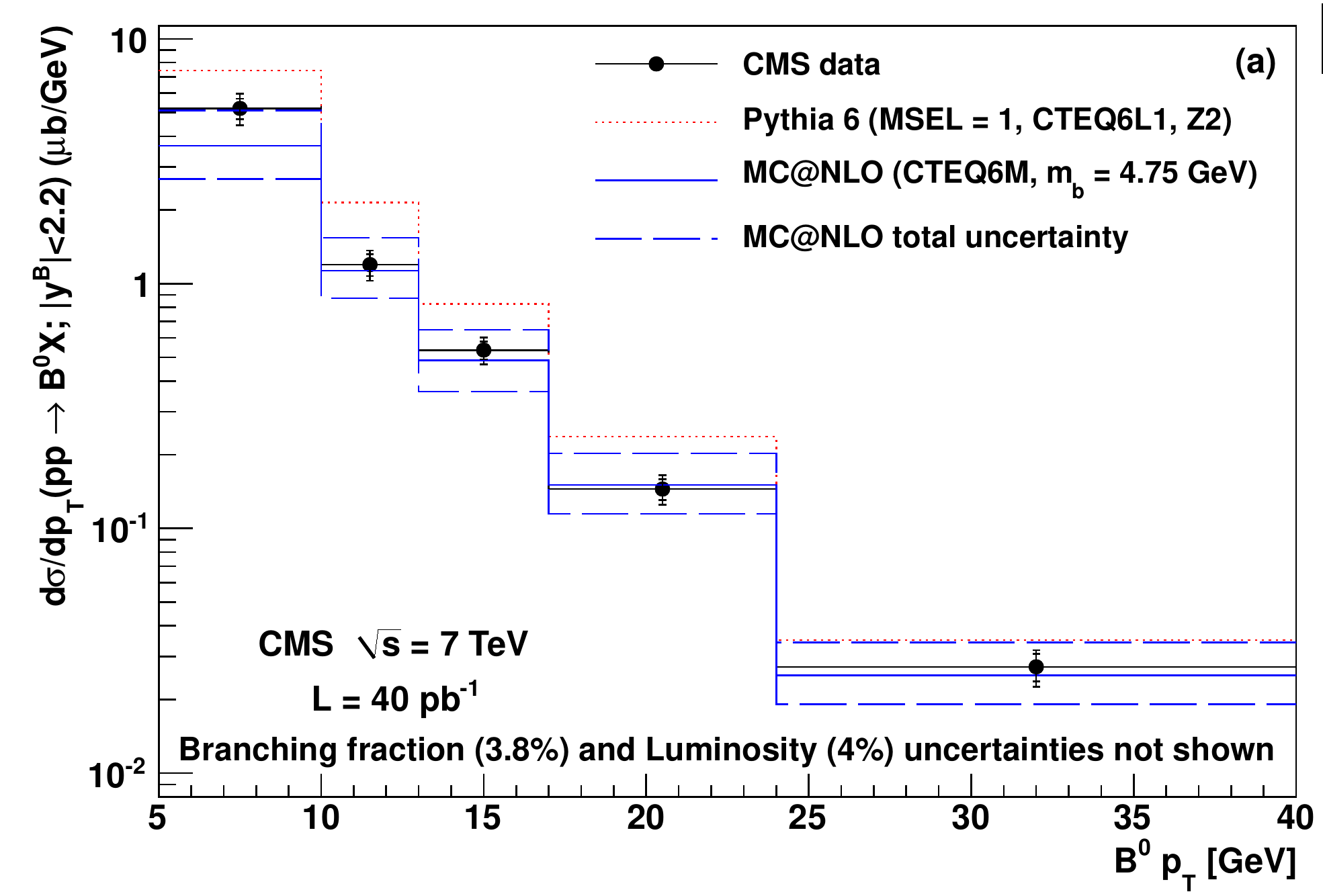}
\includegraphics[width=0.49\textwidth]{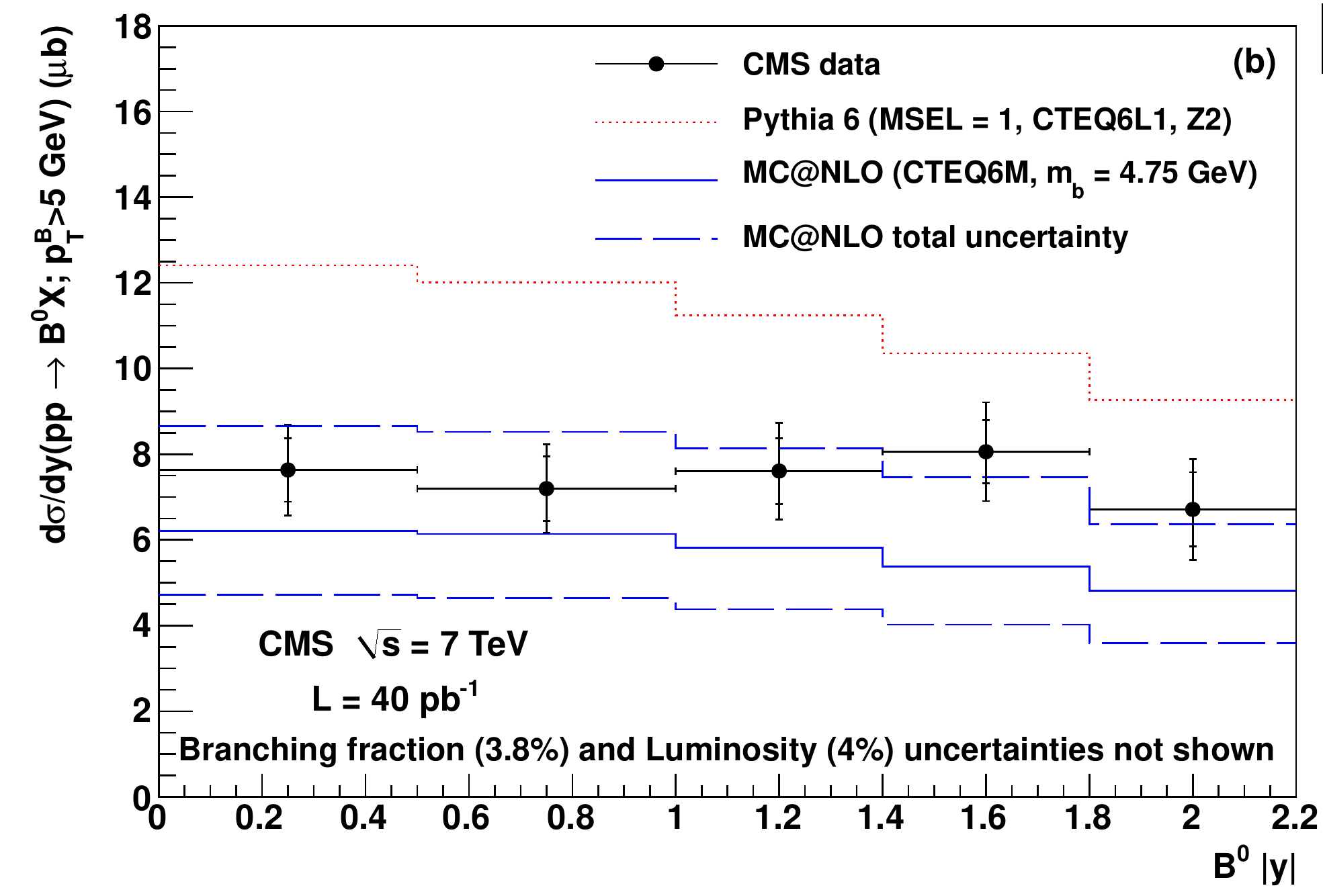}
\caption{Measured differential cross sections 
(a) $\dsdpt$ and (b) $\dsdy$ compared to the 
theoretical predictions.  The inner error bars correspond to the statistical uncertainties
and the outer error bars represents the uncorrelated
systematic uncertainties added in quadrature to the statistical uncertainties. 
Overall uncertainties of 4\% for the luminosity
and 3.8\% for the branching fractions are not shown.
The solid and dashed (blue) lines are the \MCATNLO\ prediction and its uncertainty, respectively.
The dotted (red) line is the \PYTHIA\ prediction.}
\label{fig:Xsec}
\end{center}
\end{figure}

The fitted signal yields in each bin of $\ptb$ and
$\yb$ are summarized in Table~\ref{tab:results}. 
Figure~\ref{fig:fit} shows the fit projections for $\MB$ and $ct$ from the
inclusive sample with $\ptb>5\GeV$ and $\yb < 2.2$.
The total number of signal events is $809\pm 39$, 
where the uncertainty is statistical only.

The differential cross section is calculated in bins of $\ptb$ as
\begin{equation}
\frac{d\sigma(\Pp\Pp\to \PBz X)}{dp_\mathrm{T}^{\PB}} = \frac{n_{\rm sig}}
{2\cdot\epsilon \cdot{\cal B}\cdot L \cdot\Delta \ptb},
\end{equation}
and similarly for $\yb$, where $n_\text{sig}$ is the fitted number of signal 
events in the given bin, $\epsilon$ is the efficiency for a $\PBz$ meson 
to pass all the selection criteria, $L$ is the integrated luminosity,
$\Delta \ptb$ is the bin size, and ${\cal B}$ is the product of branching
fractions ${\cal B}(\Bpsiks) = (4.36\pm 0.16)\times 10^{-4}$,
${\cal B}(\psimumu)=(5.93\pm 0.06)\times 10^{-2}$, and 
${\cal B}(\kspipi)= 0.6920\pm 0.0005$~\cite{PDG2010}.  
The additional factor of two in the denominator accounts for our choice of quoting
the cross section for $\PBz$ production only, while $n_{\rm sig}$ includes both \PBz\ and
\PaBz.  
The efficiencies are calculated separately for each bin, always considering only mesons produced with
$\yb<2.2$ ($\ptb>5\GeV$) for \ptb\ (\yb) bins,
and take into account bin-to-bin 
migrations ($<1\%$) due to the resolution on the measured \ptb\ and \yb.

\begin{table*}[!tb]
\caption{Signal yield $n_{\rm sig}$, 
efficiency $\epsilon$, and measured differential cross sections $\dsdpt$ 
and $\dsdy$, compared to the \MCATNLO~\cite{MCNLO} and \PYTHIA~\cite{PYTHIA} predictions. The uncertainties 
in the measured cross sections are statistical and systematic, respectively, excluding 
the common luminosity ($4\%$) and branching fraction ($3.8\%$) uncertainties. The uncertainties on
the signal yields are statistical only, while those on the efficiencies are systematic.
}
\label{tab:results}
\begin{center}
\begin{tabular*}{1.0\textwidth}{@{\extracolsep{\fill}}cccccc} 
\hline
\hline
$\ptb~(\!\GeV)$ \T \B      & $n_\text{sig}$ & $\epsilon~(\%)$     & $\dsdpt~(\!\ub/\!\GeV)$		& $\MCATNLO$ & $\PYTHIA$ \\ 
\hline
$5-10$  & $240\pm23$ &  $0.65\pm0.05$ & $5.20\pm0.50\pm0.59$    & 3.66 & 7.42 \\
$10-13$ & $169\pm17$ &  $3.32\pm0.28$ & $1.196\pm0.121\pm0.117$ & 1.13 & 2.14 \\
$13-17$ & $193\pm16$ &  $6.37\pm0.51$ & $0.535\pm0.045\pm0.051$ & 0.49 & 0.83 \\
$17-24$ & $138\pm13$ &  $9.60\pm0.76$ & $0.145\pm0.014\pm0.014$ & 0.15 & 0.24 \\
$24-40$ &  $70\pm9$  & $11.40\pm1.04$ & $0.027\pm0.003\pm0.003$ & 0.025 & 0.035 \\
\sgline
$|y^B|$ & $n_\text{sig}$ & $\epsilon~(\%)$ & $\dsdy~(\!\ub)$ & $\MCATNLO$ & $\PYTHIA$ \\
\sgline
$0.0-0.5$ & $145\pm14$ & $1.34\pm0.10$ & $7.63\pm0.74\pm0.76$ & 6.21 & 12.41 \\
$0.5-1.0$ & $141\pm15$ & $1.38\pm0.10$ & $7.20\pm0.75\pm0.71$ & 6.14 & 12.01 \\
$1.0-1.4$ & $167\pm17$ & $1.93\pm0.15$ & $7.61\pm0.77\pm0.83$ & 5.81 & 11.24 \\
$1.4-1.8$ & $229\pm21$ & $2.51\pm0.21$ & $8.06\pm0.74\pm0.89$ & 5.38 & 10.36 \\
$1.8-2.2$ & $128\pm17$ & $1.69\pm0.14$ & $6.71\pm0.87\pm0.80$ & 4.81 & 9.26 \\
\hline
\end{tabular*}
\end{center}
\end{table*}

The cross section is affected by systematic uncertainties on the signal yield and efficiencies, which are uncorrelated 
bin-to-bin and can affect the shapes of the distributions, and by uncertainties on the branching fractions
and luminosity, which are common to all bins and only affect the overall normalization.  
The uncertainty on the signal yield arises from potential 
fit biases and imperfect knowledge of the PDF parameters ($4$--$7\%$),
and from effects of final-state radiation and mismeasured track momenta on the signal shape 
in $\MB$ ($1\%$).  Uncertainties on the efficiencies arise from the trigger ($2$--$3\%$), 
muon identification ($1\%$), muon tracking ($1\%$), $\PKzS$ ($5\%$) and $\PBz$
($3\%$) candidate selection requirements, acceptance ($2$--$3\%$),
dimuon correlations ($1$--$5\%$) and \ptb\ and \yb\ mismeasurement ($1\%$). The first five
efficiency uncertainties are determined directly from data, while the last three are 
determined by simulation. The largest uncertainties on the efficiency arrise from the $\PKzS$
reconstruction, which is dominated by the displaced hadronic track efficiency and is measured
by comparing the reconstructed $\PKzS$ lifetime with the known value, and the dimuon correlation
uncertainty, which is taken as $100\%$ of the correction applied to account for the correlations.
The difference between the kinematically reweighted and unreweighted results
($3$--$5\%$) 
is taken as an additional systematic uncertainty.
The bin-to-bin systematic uncertainty
is computed as the sum in quadrature of the individual uncertainties, and is 
summarized in Table~\ref{tab:results}.  In addition, there are normalization
uncertainties of $4\%$ from the luminosity measurement
and of $3.8\%$ from the branching fractions~\cite{PDG2010}.

The differential cross sections as functions of $\ptb$ and $\yb$ are shown in 
Fig.~\ref{fig:Xsec} and Table~\ref{tab:results}.  They are compared to the predictions of 
\MCATNLO~\cite{MCNLO} using a \cPqb-quark mass $m_{\cPqb}$ of $4.75\GeV$, renormalization and factorization scales
$\mu = \sqrt{m_{\cPqb}^2+\pt^2}$, and the CTEQ6M parton distribution 
functions~\cite{CTEQ}.  The uncertainty on the predicted cross section is calculated by 
independently varying the 
renormalization and factorization scales by factors of two, $m_{\cPqb}$ by $\pm 0.25\gev$, and by
using the CTEQ6.6 parton distribution functions.
For reference, the prediction of \PYTHIA~\cite{PYTHIA} is also included, using a \cPqb-quark mass of 
$4.80\GeV$, CTEQ6L1 parton distribution functions~\cite{CTEQ}, and the Z2 tune~\cite{Z2} to simulate the 
underlying event. The measured \pt\ spectrum falls slightly faster than predicted by \MCATNLO, 
while the $y$ spectrum is measured to be flatter than the
\PYTHIA\ prediction and in agreement with the \MCATNLO\ prediction within
uncertainties. The integrated cross section for $\ptb > 5\GeV$ and $\yb < 2.2$ is 
calculated as the sum over all \pt\ bins, without an upper limit for the highest \pt\ bin,
to be \totalsigma, where the first uncertainty is statistical and the 
second is systematic.  The result is compatible with the 
prediction from \MCATNLO\ ($25.2^{+9.6}_{-6.2}~\mu$b) and below
the prediction from $\PYTHIA$ ($49.1~\mu$b).

In summary, the first measurements of the differential cross sections $\dsdpt$ and $\dsdy$ for 
$\PBz$ mesons produced in $\Pp\Pp$ collisions at $\sqrt{s}=7\TeV$ have been presented 
using the decay $\Bpsiks$.
The measurements cover a range in $\ptb$ from $5\GeV$ to more than $30\GeV$, 
and the rapidity range $\yb < 2.2$.  The total cross section in this kinematic region
lies between 
the central values of the \MCATNLO\ and \PYTHIA\ predictions, with a rapidity distribution that is 
flatter than \PYTHIA. It is also in agreement within uncertainties with the 
measured $\PB^+$ cross section~\cite{BPH-10-004}.

We wish to congratulate our colleagues in the CERN accelerator departments for the
excellent performance of the LHC machine. We thank the technical and administrative
staff at CERN and other CMS institutes, and acknowledge support from: FMSR (Austria);
FNRS and FWO (Belgium); CNPq, CAPES, FAPERJ, and FAPESP (Brazil); MES (Bulgaria); CERN;
CAS, MoST, and NSFC (China); COLCIENCIAS (Colombia); MSES (Croatia); RPF (Cyprus);
Academy of Sciences and NICPB (Estonia); Academy of Finland, ME, and HIP (Finland);
CEA and CNRS/IN2P3 (France); BMBF, DFG, and HGF (Germany); GSRT (Greece);
OTKA and NKTH (Hungary); DAE and DST (India); IPM (Iran); SFI (Ireland); INFN (Italy);
NRF and WCU (Korea); LAS (Lithuania); CINVESTAV, CONACYT, SEP, and UASLP-FAI (Mexico);
PAEC (Pakistan); SCSR (Poland); FCT (Portugal); JINR (Armenia, Belarus, Georgia,
Ukraine, Uzbekistan); MST and MAE (Russia); MSTD (Serbia); MICINN and CPAN (Spain);
Swiss Funding Agencies (Switzerland); NSC (Taipei); TUBITAK and TAEK (Turkey);
STFC (United Kingdom); DOE and NSF (USA).

\bibliography{auto_generated}   

\cleardoublepage \section{The CMS Collaboration \label{app:collab}}\begin{sloppypar}\hyphenpenalty=5000\widowpenalty=500\clubpenalty=5000\textbf{Yerevan Physics Institute,  Yerevan,  Armenia}\\*[0pt]
S.~Chatrchyan, V.~Khachatryan, A.M.~Sirunyan, A.~Tumasyan
\vskip\cmsinstskip
\textbf{Institut f\"{u}r Hochenergiephysik der OeAW,  Wien,  Austria}\\*[0pt]
W.~Adam, T.~Bergauer, M.~Dragicevic, J.~Er\"{o}, C.~Fabjan, M.~Friedl, R.~Fr\"{u}hwirth, V.M.~Ghete, J.~Hammer\cmsAuthorMark{1}, S.~H\"{a}nsel, M.~Hoch, N.~H\"{o}rmann, J.~Hrubec, M.~Jeitler, G.~Kasieczka, W.~Kiesenhofer, M.~Krammer, D.~Liko, I.~Mikulec, M.~Pernicka, H.~Rohringer, R.~Sch\"{o}fbeck, J.~Strauss, F.~Teischinger, P.~Wagner, W.~Waltenberger, G.~Walzel, E.~Widl, C.-E.~Wulz
% ...
\vskip\cmsinstskip
% ...
1:~~Also at CERN, European Organization for Nuclear Research, Geneva, Switzerland\\
% ...

\end{sloppypar}
\end{document}